\documentclass[12pt,a4paper]{article}

\usepackage{mathrsfs}
\usepackage[margin=2.5cm]{geometry}
\usepackage{amsmath,amssymb}
\usepackage[hidelinks]{hyperref}
\hypersetup{pdfauthor={Victor Ovidiu Șlupic}}
\let\originalnewline\newline
\renewcommand{\newline}{\originalnewline\hspace*{\parindent}}

\newcommand{\defe}{\overset{\mathrm{def}}{=}} 

\begin{document}

\hypersetup{pageanchor=false}

\begin{titlepage}
\vspace*{3cm}

\begin{center}
{\Large\bfseries
The Feynman Dyson argument extended to the inhomogeneous Maxwell equations\par}

\vspace{0.5cm}

Victor Ovidiu Șlupic\textsuperscript{1,2}

\vspace{0.2cm}

{\small\itshape
\textsuperscript{1}Horia Hulubei National Institute for R\&D
in Physics and Nuclear Engineering,\\
Department of Theoretical Physics,\\
Reactorului 30, Bucharest-Magurele, 077125, Romania\\[0.2cm]
\textsuperscript{2}University of Bucharest,\\
Faculty of Physics,\\
Atomiștilor 405, Măgurele, 077125, Romania\par}

\vspace{0.2cm}

{\small
\href{mailto:victor-ovidiu.slupic@s.unibuc.ro}
{\texttt{victor-ovidiu.slupic@s.unibuc.ro}}\par}

\vspace{0.5cm}

\end{center}

\vspace{1cm}

\begin{abstract}The Feynman-Dyson argument gives the homogeneous Maxwell equations but does not identify the sources in the other two equations with the charge and current carried by matter. Taking the homogeneous equations as given, we derive Gauss’s law and the Ampère-Maxwell law for the sources entering the Lorentz force and the power delivered to matter. We assume, in particular, that the field densities and fluxes are at most quadratic in the fields and covariant under proper rotations. We require the same expressions for the sources to satisfy local energy and momentum conservation for every field that obeys the homogeneous equations, and to be continuous in spacetime along every smooth such field. The charge and current densities are then unique up to the values of $\varepsilon_0$ and $\mu_0$. We also discuss how these assumptions might be relaxed.
\end{abstract}

\end{titlepage}
\hypersetup{pageanchor=true}

\tableofcontents

\section{Introduction}

In his 1865 account of the electromagnetic field, Maxwell brought electrical, magnetic and optical phenomena into a common dynamical framework. His theory connected the propagation of light with electromagnetic disturbances and related the energy of the field to the mechanical forces acting on electrified bodies and currents \cite{Maxwell1865}. These connections invite a further investigation of the assumptions underlying the theory, particularly the relation between the action of a field on matter and the role of matter as its source.\newline
In October 1948, Feynman showed Dyson an argument that approached electrodynamics from the motion of a single nonrelativistic particle. Starting from Newton's equation  $F_j(x,\dot{x},t) =m \ddot{x}_j$ and the commutation relations $[x_i, x_j]=0, \quad m[x_i,\dot{x}_j] = i \hbar \delta_{ij}$, he obtained the Lorentz form of the
force and the two homogeneous Maxwell equations. According to Dyson, Feynman sought to explore particle dynamics without initially assuming the usual Hamiltonian or Lagrangian formulation. The argument remained unpublished until 1990, when Dyson presented a reconstruction of Feynman's exposition together with an account of its motivation \cite{Dyson1990}.\newline
The equations containing charge and current occupy a different position in Dyson's presentation. He remarks that they ``merely define'' the external charge and current densities \cite{Dyson1990}. This statement leaves a physical
identification unresolved. Given electric and magnetic fields, one can introduce quantities through the differential expressions appearing in Gauss's law and the Ampère-Maxwell law. Such definitions do not, however, establish that these quantities coincide with the charge and current carried by matter. Once $\rho$ and $J$ denote the physical densities entering the force and power delivered to matter, the source equations express physical laws relating the electromagnetic field to the charge and current carried by matter. Their validity requires an argument beyond the introduction of notation. Subsequent work has examined the compatibility of Galilean mechanics with the full Maxwell equations \cite{VaidyaFarina1991} and the role of minimal coupling, where a conserved current is introduced through the divergence of the field tensor \cite{Montesinos1999}. Harikumar et al. extend the minimal coupling construct to $\kappa$ Minkowski spacetime to first order in the deformation, defining a deformed conserved current \cite{Harikumar2011}, while Kauffman develops a noncommutative generalization of the Feynman-Dyson argument \cite{Kauffman2022}. Di Mauro et al. reconstruct Feynman's later route to the source equations from electrostatics, relativistic covariance and charge conservation \cite{DiMauro2021}.\newline
We take this unresolved identification as the starting point of the present work. Retaining the homogeneous equations obtained in the Feynman-Dyson argument, we investigate which additional assumptions allow the source equations to follow from local conservation of energy and momentum. The physical motivation is direct: the force exerted on matter and the work done on it must be balanced by changes and transport of field momentum and energy. The same charge and
current must enter both descriptions of this exchange. We therefore introduce $\rho$ and $ J$ through their role in the Lorentz force and electrical power, and seek to determine their relation to the fields. \newline
The use of conservation principles in reconstructing electrodynamics has substantial antecedents. Imai developed a formulation based on energy and momentum conservation \cite{Imai1991}, while N\"ockel obtained Maxwell's equations from local energy and momentum conservation within relativistic mechanics, using independent variation of the field values at fixed derivatives \cite{Nockel2022}. Assuming the Maxwell energy and momentum tensor in a relativistic setting, Arminjon also recovered the source equations from the conservation laws and the Lorentz force where $E \cdot B \neq 0$ \cite{Arminjon2016,Arminjon2018}. Related work by Navarro and Sancho characterizes electromagnetic energy tensors through geometric and divergence conditions \cite{Navarro2012}. In macroscopic electrodynamics, Gratus, McCall and Kinsler expressed free charge and current through first order linear differential operators acting on $E$ and $B$, eliminating the excitation fields $D$ and $H$ \cite{Gratus2020}. Complementary classifications determine local conservation laws assuming the full source free Maxwell equations \cite{AncoPohjanpelto2001}. Our investigation concerns the particular extension needed by the Feynman-Dyson argument: the homogeneous equations are given, while the source equations and the admissible field densities and fluxes are to be determined without postulating covariance under boosts.\newline
To make this reconstruction precise, we assume that the field energy, momentum, energy flux and stress depend algebraically on the electric and magnetic fields at the same point. Their dependence is restricted
to polynomials of degree at most two, with constant coefficients and covariance under proper spatial rotations. We assume that the charge and current densities are finite functions of the field values and first derivatives at the same point. The same functions must satisfy both conservation laws for every choice of these data compatible with the homogeneous equations. This is an additional assumption i.e. the same source functions must satisfy both conservation laws for every field configuration allowed by the homogeneous equations, not just for one particular configuration. Along every smooth field satisfying those equations, the assigned sources must be continuous in spacetime.\newline
Under these assumptions, we obtain Gauss's law and the Ampère-Maxwell law for the physical sources entering mechanical exchange. For fixed electromagnetic constants, the stated assumptions uniquely determine the charge and current densities from the fields and their first derivatives. The same analysis determines the field densities and
fluxes up to two explicitly conserved additions and constant offsets. After subtracting the constant energy offset, positivity of the field energy selects positive values of the two electromagnetic constants. Neither parity, time reversal, covariance under boosts nor a field action is assumed in the reconstruction. \newline
We first state the premises and classify the permitted field quantities, then derive the source equations and examine the freedom left by conservation. Particular attention is given to configurations in which the electric and magnetic fields have vanishing scalar product, where continuity is needed to complete the source identification. Finally, a nonlinear example in the appendix shows that the other premises do not imply the quadratic restriction. The result is therefore a conditional extension of the Feynman-Dyson argument, with the assumptions responsible for recovering the physical source equations stated explicitly.
\section{Physical postulates and algebraic freedom}
In this paper we shall not assume special relativity. Latin indices take values from $1$ to $3$, with repeated indices summed. The fields have continuous first derivatives and satisfy
\begin{equation}\label{maxwell12}
\partial_a B_a = 0, \quad \partial_t B_a = - \varepsilon_{abc} \partial_b E_c.
\end{equation}
We furthermore assume that the physical charge and current densities respectively $\rho$ and $J$ that enter the Lorentz force, also enter the local conservation laws of momentum and energy.
\begin{equation}\label{momentum}
\partial_{b}T_{ab} - \partial_t g_a = \rho E_a + \varepsilon_{abc} J_b B_c,
\end{equation}
\begin{equation}\label{energy}
\partial_t u +\partial_a S_a = - E_a J_a.
\end{equation}

Here $-T_{ab}$ is the outward flux of momentum component $a$ across a surface normal to direction $b$, $u$ and $g$ are the field energy density and field momentum density and $S$ is the outward energy flux. No relation between momentum and velocity is assumed for matter. We also exclude singular point-charge self fields. \newline
The four field quantities have an algebraic dependence on $E$ and $B$ at the same point, with constant coefficients. No derivatives or matter variables enter in their definition. We assume that these expressions are polynomials of total degree at most two in the field components and satisfy proper rotational covariance which permits us to write 

\begin{align}\label{algebraic-family}
T_{ab} = &w_1 E_a E_b + w_2 B_a B_b + \delta_{ab} (w_3 E^2 + w_4 B^2) \\ \nonumber
&+w_5 E_aB_b + w_6 B_a E_b + w_7 \delta_{ab} E \cdot B \\ \nonumber
&+h_E \varepsilon_{abc} E_c + h_B \varepsilon_{abc}B_c \\ \nonumber
g = &a_E E + a_B B + r E \times B \\ \nonumber
S = &d_E E + d_B B + s E \times B \\ \nonumber
u = &z_E E^2 + z_B B^2 + z_C E \cdot B.
\end{align}
There are eighteen real coefficients, constant contributions to $u$ and a constant isotropic contribution to $T$ have been subtracted. Since rotations do not distinguish between polar and axial vectors, we shall keep the linear and mixed structure in \eqref{algebraic-family}. In particular, the stress tensor $T_{ab}$ has not been assumed symmetric.
After eliminating $\partial_t B$ by using \eqref{maxwell12}, the data at a point are
\begin{equation}\label{data}
D \defe (E_a, B_a, \partial_b E_a, \partial_b B_a, \partial_t E_a) , \quad \partial_a B_a =0.
\end{equation}
These data form a space of dimension $26$. The derivatives and field values can be prescribed independently subject only to $\partial_a B_a =0$. The affine fields
\begin{align}\label{affine}
&\nonumber E_a (x,t) = e_a + A_{ab} x_b + v_a t, \\ 
&B_a(x,t) = b_a + C_{ab} x_b - \varepsilon_{abc}A_{cb}t, \quad C_{aa} =0,
\end{align}
realize any such data and satisfy \eqref{maxwell12}.
We further assume that the physical sources are determined by a single pair of functions i.e. $\rho(D)$ and $J(D)$, taking finite real values at every admissible $D$. These functions must satisfy the momentum and energy conservation laws, \eqref{momentum} and \eqref{energy}, for every such $D$, with the coefficients in \eqref{algebraic-family} held fixed. Moreover, for every smooth field pair satisfying \eqref{maxwell12}, the resulting sources $\rho(D(x,t))$ and $J(D(x,t))$ must be continuous in spacetime. No separate assumption of continuity on the full data space is imposed.
These requirements apply to every smooth field pair satisfying \eqref{maxwell12}, irrespective of whether it arises from a matter system. The affine construction above establishes that every admissible $D$ can be realized by fields satisfying the homogeneous equations. It does not establish that those data can be realized by a particular matter system. Compatibility with the conservation laws throughout the admissible data space is an additional postulate.
Field energy positivity $u(E,B) > 0$ for every field pair $(E,B) \neq (0,0)$ will be used to select the signs of the two constants remaining after the reconstruction. It plays no role in the coefficient elimination or the identification of the source expressions.

\section{Deriving the source equations}
\subsection{Algebraic restrictions from the conservation laws}
Let us define the left hand side of momentum and energy conservation relations as the following field quantities
\begin{equation}\label{fields}
R_a \defe \partial_b T_{ab} - \partial_t g_a, \quad Q \defe \partial_t u + \partial_a S_a.
\end{equation}
At $E=B=0,$ finite sources have vanishing force and power. The terms in \eqref{algebraic-family} that are linear in the fields therefore give 
\begin{align}\label{zero}
& \nonumber 0 = (h_E +a_B) \nabla \times E + h_B \nabla \times B - a_E \partial_t E, \\
& 0 = d_E \nabla \cdot E.
\end{align}
The derivative data are independent, thus
\begin{equation}\label{linear}
a_E = h_B = d_E = 0, \quad h_E = - a_B.
\end{equation}
The remaining terms proportional to $a_B$ and $d_B$ satisfy the following homogeneous conservation identities
\begin{align}\label{additions}
\partial_b(-a_B \varepsilon_{abn} E_n) & - \partial_t (a_B B_a) =0, \nonumber \\
& \; \; \; \nabla \cdot (d_B B) =0.
\end{align}

Therefore they can consequently be removed from $R$ and $Q$ without imposing further conditions on the sources.
For perpendicular non-zero fields, conservation requires

\begin{equation}\label{compat}
B \cdot R = 0, \quad B^2 Q + (E \times B) \cdot R = 0.
\end{equation}
These are necessary conditions for finite sources, obtained from \eqref{momentum} and \eqref{energy}, without division by $E \cdot B$. 
Let us now choose $E = E_0 \hat{e}_1$ and $B = B_0 \hat{e}_2$, where by $\hat{e}_a$ we understand a Cartesian unit vector. 
In the first identity \eqref{compat}, the coefficients of $\partial_1 E_1, \partial_2 B_1$ and $\partial_2 E_2$ are respectively
\begin{equation}
w_6 B_0^2, \quad w_7 E_0 B_0, \quad (w_5+w_6+w_7)B_0^2.
\end{equation}

They must vanish separately. The remaining terms then require

\begin{equation}
0=(w_1 -r)E_0 B_0 \partial_1 E_2 + (2w_3 +r)E_0 B_0 \partial_2 E_1 + (w_2 + 2w_4)B_0^2 \partial_2 B_2.
\end{equation}
The constraint on $\nabla \cdot B $ is fulfilled by setting $\partial_1 B_1=0$  and $\partial_3 B_3 = - \partial_2 B_2$. Thus

\begin{equation}\label{coefficients}
w_5 = w_6 = w_7 =0, \quad w_1 = r, \quad w_3 = - \frac{w_1}{2}, \quad w_4 = - \frac{w_2}{2}.
\end{equation}

We define
\begin{equation}
L \defe w_2 \nabla \times B - w_1 \partial_t E.
\end{equation}

Differentiation of \eqref{algebraic-family}, using \eqref{linear} and \eqref{coefficients} yields
\begin{align}\label{expanded}
&R = w_1 E \nabla \cdot E + L \times B, \nonumber \\
&Q = (s-2 z_B)B \cdot \nabla \times E - s E \cdot \nabla \times B + 2 z_E E \cdot \partial_t E + z_C (B\cdot \partial_t E - E \cdot \nabla \times E).
\end{align}

In the second identity of \eqref{compat}, the coefficient of $\partial_t E_2$ is $z_C B_0^3$, so $z_C = 0,$ and the identity reduces to

\begin{equation}
0 = (s-w_2) E_0 (\partial_2 B_3 - \partial_3 B_2) + (2 z_B - s) B_0 (\partial_3 E_1 - \partial_1 E_3) +(w_1 - 2 z_E) E_0 \partial_t E_1.
\end{equation}
Independent variation of the three derivative combinations gives
\begin{equation}
s=w_2, \quad z_E = \frac{w_1}{2}, \quad z_B = \frac{w_2}{2}, \quad z_C = 0.
\end{equation}

Neither parity nor time reversal has entered this argument.
\subsection{Identification of the physical sources}
With the coefficients fixed, the conservation equations become 
\begin{align}\label{differences}
&(\rho - w_1 \nabla \cdot E) E + (J-L) \times B =0, \nonumber \\ 
& E \cdot (J-L) =0.
\end{align}
Where $E \cdot B \neq 0$ and contraction of the first equation with $B$ determines $\rho$. The vector equation gives $J-L = \lambda B$. The energy equation gives $\lambda E \cdot B =0$, hence $\lambda =0$. Consequently,
\begin{equation}\label{sources}
\rho = w_1 \nabla \cdot E, \quad J = w_2 \nabla \times B - w_1 \partial_t E.
\end{equation}

These equalities identify the sources entering through mechanical exchange, and are not introduced as definitions of $\rho$ and $J$ respectively.
To extend \eqref{sources} such that we have compatibility with a given datum $D_0$ with $E \cdot B=0$, we shall take the affine fields \eqref{affine} and add a perturbation of the form
\begin{equation}\label{perturbation}
\delta E = E_0 \bigg( \frac{t}{\tau_0} \bigg)^2 \hat{e}_2, \quad \delta B = B_0 \bigg( \frac{x_1}{l_0}\bigg)^2 \hat{e}_2,
\end{equation}
where $E_0,B_0 \neq 0$ have dimensions of fields and $\tau_0, l_0>0$ are arbitrary constants with dimensions of time and length. Both perturbations have zero values and zero first derivatives at the origin. They preserve \eqref{maxwell12} since $\nabla \cdot \delta B  =0,\nabla \times \delta E =0, \partial_t \delta B= 0$. The resulting fields thus realize the same datum $D_0$.
Along $x_1 = l_0 s, x_2=x_3=0, t= \tau_0 s$, their scalar product i.e. $E' \cdot B' =(E+\delta E) \cdot (B+ \delta B)$ is a polynomial in the dimensionless parameter $s$. Its coefficient of $s^4$ is $E_0 B_0 \neq 0$ and all other contributions have degree at most three. It is thus nonzero for all sufficiently small nonzero $s$. Equation \eqref{sources} holds there. Spacetime continuity of the assigned sources and of the field derivatives extends it to vanishing $s$, establishing the source laws for arbitrary datum $D_0$.\newline
This argument is used to replace an independent data-space continuity postulate by spacetime continuity for every smooth admissible field. It does not remove the universal quantifier i.e. spacetime continuity is still required for every smooth field satisfying the homogeneous equations. An alternative sufficient formulation is continuity in field values at fixed derivatives at $E\cdot B = 0:$ perpendicular fields can be approached by non perpendicular ones with derivatives unchanged, and \eqref{sources} then extends by the same limiting argument. We note that neither formulation is asserted to be logically minimal among all possible source assumptions and further directions to modifying the assumptions are discussed in section five.

\subsection{Field energy, momentum and fluxes}
Since the energy $u$ must be positive, we impose $w_1, w_2 >0$. We can now write
\begin{equation}
\varepsilon_0 \defe w_1, \quad \mu_0^{-1} \defe w_2, \quad c^2 \defe \frac{1}{\varepsilon_0 \mu_0}.
\end{equation}

The complete family compatible with our assumptions is 

\begin{gather}
 u = \frac{ \varepsilon_0 E^2}{2} + \frac{B^2}{2 \mu_0}, \nonumber \\
 g = \varepsilon_0 E \times B + \alpha B, \nonumber \\
  S = \mu_0^{-1} E\times B + \beta B, \nonumber \\ 
 T_{ab} = \varepsilon_0 \bigg(E_a E_b - \frac{1}{2} \delta_{ab} E^2\bigg) + \mu_0^{-1} \bigg( B_a B_b - \frac{1}{2} \delta_{ab} B^2 \bigg) - \alpha \varepsilon_{abc} E_c,
 \label{densities}
\end{gather}

where $\alpha \defe a_B$ and $\beta \defe d_B$ are arbitrary dimensionful real constants. Together with \eqref{sources}, these expressions satisfy both conservation laws for all admissible data. This can be verified using equations \eqref{expanded} and \eqref{additions}. \newline
The source laws are unique, whereas the densities and fluxes retain additions that satisfy homogeneous conservation identities. Hehl and Obukhov discuss the freedom to add a closed current to the electromagnetic energy and momentum current in a premetric formulation \cite{HehlObukhov2001}. Arminjon's polynomial uniqueness result assumes that two energy currents have the same divergence for arbitrary independent field values and first derivatives \cite{Arminjon2016energy}. Here the homogeneous Maxwell equations constrain those derivatives. Selecting $\alpha = \beta =0$ gives the conventional Maxwell equations densities and fluxes, for which $g = \frac{S}{c^2}$ and the stress tensor $T_{ab}$ is symmetric. The relation $g = \frac{S}{c^2}$ must not be attributed to conservation alone. \newline
For twice continuously differentiable fields, \eqref{sources} implies charge conservation and in a source free region it combines with \eqref{maxwell12} to give wave equations with speed $c$. We note also that the resulting field equations also admit their conventional Lorentz covariant formulation, of which we make no use. This does not establish a boost transformation law for matter, a universal limiting speed, or a coupled matter theory. \newline
If strict positivity of the normalized energy is relaxed to nonnegativity, the coefficient family also includes two degenerate cases. With $\alpha = \beta = 0$, the choice $w_1>0$, $w_2=0$ gives the electric Carroll equations with $S=0$ \cite{Duval2014,deBoer2022}, whereas $w_1 =0$, $w_2 >0$ gives a Galilei invariant magnetic sector with $\rho =0$ and $g=0$ \cite{Duval2014}. In both cases the energy vanishes for some nonzero field pairs, so neither satisfies the strict positivity assumed here.

\section{Residual freedom}
\subsection{Finiteness and continuity of the sources}
Define $\Delta \rho \defe \rho - \varepsilon_0 \nabla \cdot E$ and $\Delta J \defe J -L$. Fix a spacetime point and hold the field values and all their first derivatives fixed there. This is the meaning of ``at that point'' below. \newline
 If $B \neq 0$ and $E\cdot B =0$ all solutions of \eqref{differences} are
\begin{equation}
\Delta J = \Delta \rho \frac{E \times B}{B^2} + \lambda B,
\end{equation}
where $\Delta \rho$ and $\lambda$ may be any finite real numbers. Nonzero choices of either $\Delta \rho$ or $\lambda$ can change the source values while preserving momentum and energy conservation \eqref{momentum} and \eqref{energy}, at that point. The source differences obey the same algebraic kernel condition as the current in force free electrodynamics, where Maxwell's source equation is assumed \cite{GrallaJacobson2014,GrallaJacobson2024Correction}.\newline
If $B =0, E \neq 0,$ the allowed differences satisfy $\Delta \rho =0$ and $E \cdot \Delta J =0$. \newline
If $ E = B = 0$, both differences are unconstrained at the point. Energy and momentum conservation with finite sources give the compatibility conditions \eqref{zero} and \eqref{compat} used to fix the algebraic coefficients. They do not, at a single datum, eliminate the source differences. To meet the full premises, the same source functions must satisfy the conservation laws at every admissible datum, with $\rho$ and $J$ continuous in spacetime for every smooth field satisfying the homogeneous Maxwell equations. \newline
The requirement of spacetime continuity cannot be replaced by local boundedness of the source functions. Even if these functions remain bounded throughout a neighbourhood of each admissible datum, they may satisfy both conservation laws at every datum while differing from the Maxwell source laws. For a nonzero constant $k$ of the appropriate dimensions, set $\Delta \rho =0$ and 

\begin{equation}
\Delta J =
\begin{cases}
k B, & E \cdot B = 0,\\
0, & E \cdot B \neq 0.
\end{cases}
\end{equation}
This assignment is locally bounded, since $| \Delta J | \leq |k| |B|$, and satisfies \eqref{differences} at every datum. Yet it differs from \eqref{sources} whenever $E \cdot B = 0$ and $B \neq 0$.
This assignment does, however, violate the required spacetime
continuity. To see this, fix an admissible datum $D_0$ with
$E \cdot B = 0$ and $B \neq 0$. Realize it at the origin
using the affine fields \eqref{affine}, and add the perturbations
\eqref{perturbation}, which leave $D_0$ unchanged.
Along the curve $x_1=l_0s$, $x_2=x_3=0$, $t=\tau_0s$,
the resulting fields satisfy $E' \cdot B' \neq 0$ for all
sufficiently small nonzero $s$, as established above.
The assigned difference $\Delta J$ therefore vanishes along
the approach to the origin, whereas at the origin it has
the nonzero value $k B(D_0)$. Thus $\Delta J$ is discontinuous
there. Since $L$ depends continuously on the field derivatives,
the physical current $J=L+\Delta J$ is also discontinuous.
This violates the requirement that the assigned sources be
continuous in spacetime for every smooth field satisfying
the homogeneous Maxwell equations. \newline
Continuity on one chosen field configuration is also insufficient. Consider the static fields

\begin{equation}
E = 0, \qquad
B = B_0 \begin{pmatrix}
\cos(kx_3)\\
\sin(kx_3)\\
0
\end{pmatrix}.
\end{equation}
with constant $B_0 \neq 0$ and inverse length $k$. Here $\nabla \cdot B =0$, $\nabla \times B = - k B$ and $B^2$ is constant. This is a linear force free magnetic field \cite{Woltjer1958}. The Maxwell field quantities have zero stress divergence and constant energy i.e. $\partial_b T_{ab} =0, \partial_t u =0, \nabla u =0$. Every choice of $\rho = 0, J = \lambda B$, with constant $\lambda$, is smooth and satisfies momentum and energy conservation \eqref{momentum} \eqref{energy}: the Lorentz force and power densities vanish. Charge conservation also holds, since $\nabla \cdot J =0$. Nevertheless, \eqref{sources} selects only $\lambda = - \frac{k}{\mu_0}$. Compatibility and continuity on this one configuration cannot replace compatibility at every admissible datum and continuity along every smooth admissible field.

\section{Conclusion and further directions}

In this paper we have obtained the inhomogeneous Maxwell equations from the homogeneous equations and the local conservation laws of momentum and energy. We have identified the physical charge and current densities and shown that, for fixed values of the two remaining constants, their expressions are unique under the stated assumptions.

The field densities and fluxes are determined only up to terms conserved whenever the homogeneous Maxwell equations hold. The symmetry of the stress tensor and the relation $g=S/c^2$ are consequences of the choice $\alpha=\beta=0$, which gives the conventional Maxwell field quantities.

Let us summarize the assumptions. We take continuously differentiable ($C^1$) fields satisfying the homogeneous Maxwell equations and require local momentum and energy conservation with the same charge and current densities that enter the Lorentz force and power densities. We assume that the four field quantities depend only on the field values at the same point, that their expressions are polynomials of total degree at most two with constant real coefficients, and that they satisfy proper rotational covariance. We require the sources to be given by the same two functions of the fields and their first derivatives, taking finite values for every field configuration allowed by the homogeneous equations. Both conservation laws must hold along every smooth such field with the algebraic coefficients held fixed. We furthermore require the sources to remain continuous in spacetime for all such fields. Positive field energy selects the signs of the two remaining constants.

Another path would be to relax the quadratic restriction and examine which nonlinear field equations remain compatible with the same conservation laws.

The assumptions of locality could also be reconsidered by starting from interactions between separated charges. Maxwell discusses several such approaches, including those of Weber, Riemann and C. Neumann \cite{Maxwell1873Treatise,Weber1846,Riemann1867,Neumann1868}. His examination of energy conservation in Weber's theory is particularly relevant here: conservation can be investigated directly for an interacting particle system before introducing local field densities and fluxes. This suggests asking whether conservation laws for such a system, together with suitable restrictions on the interaction, can yield a local field description satisfying the Maxwell equations.

In the later theories of Schwarzschild \cite{Schwarzschild1903}, Tetrode \cite{Tetrode1922} and Fokker \cite{Fokker1929}, developed further by Wheeler and Feynman \cite{WheelerFeynman1945,WheelerFeynman1949}, charged particles interact without independent electromagnetic field degrees of freedom. In the formulation of Wheeler and Feynman, each particle responds to one half of the retarded plus one half of the advanced electromagnetic field of every other particle. The interactions thus connect events on the particle worldlines along past and future light cones. In their absorber calculation, surrounding charges absorb the emitted radiation, and their response supplies the radiation reaction. Complete absorption alone does not select retarded radiation. Wheeler and Feynman also invoke asymmetric initial conditions for the absorbing matter \cite{WheelerFeynman1945}.

For the question considered here, the difficulty is that the interaction law of Wheeler and Feynman already specifies electromagnetic propagation. Similarly, writing a direct interaction in terms of a Maxwell Green function incorporates the propagation law into the starting assumptions \cite{AndersonSchiminovich1967}. To obtain Maxwell electrodynamics from fewer assumptions by this route, one would first need physical conditions that determine the particle interaction without prescribing the Maxwell propagator. One would then need to establish when this particle theory admits a local Maxwell field description and which further conditions select its retarded radiative solutions.

\section*{Acknowledgements}
I thank Ciprian-Sorin Acatrinei for bringing this problem to my attention. I used Chatgpt $6$ Astra Ultra for proofreading and Claude Opus $5.5$ together with Undermind to suggest bibliographical references.

\appendix
\section{A nonlinear example beyond the quadratic restriction}

This appendix assumes a Lagrangian formulation of the electromagnetic field to construct a nonlinear alternative \cite{Neves2021}. A classical example of this approach is the nonlinear theory of Born and Infeld \cite{BornInfeld1934}. We use a local field Lagrangian density with the usual coupling to charge and current. This assumption is used only to construct this example which shows that the other premises do not imply the assumption that the field densities and fluxes are at most quadratic in the fields.

\subsection{Lagrangian formulation and field quantities}

Throughout this appendix, we use Heaviside-Lorentz units with the weak-field electric and magnetic coefficients, and their corresponding wave speed, equal to unity. \newline
The symbol $\mathscr{L}(X)$ denotes the field Lagrangian density. The interaction with charge and current is written separately. Choose
\begin{equation}\label{lagrangian}
X \defe \frac{E^2 -B^2}{2}, \quad \mathscr{L}(X) \defe  X +  a X^3, \quad a>0.
\end{equation}
Here $a$ has dimension $X^{-2}$, and 
\begin{equation}
\mathscr{L}'(X) \defe \frac{d\mathscr{L}}{dX}
= 1 + 3aX^2.
\end{equation}

The standard Maxwell choice in these units is $\mathscr{L}(X) =X.$ \newline
Equation $\eqref{lagrangian}$ modifies this field contribution by adding the nonlinear term. \newline
Locally, introduce the scalar potential $\phi$ and vector potential $A$ through 
\begin{equation}
E = - \nabla \phi - \partial_t A, \quad B = \nabla \times A.
\end{equation}We use the total Lagrangian density
$\mathscr{L}(X)-\rho\phi+J\cdot A$.
Define the electric displacement $D_{\rm e}$ and
magnetic intensity $H$ by
\begin{equation}
D_{\rm e} \defe \frac{\partial\mathscr{L}}{\partial E}
= \mathscr{L}'(X)E,
\qquad
H \defe -\frac{\partial\mathscr{L}}{\partial B}
= \mathscr{L}'(X)B.
\end{equation}
The Euler--Lagrange equations obtained by varying
$\phi,A$, with the sources $\rho,J$ held fixed, are
\begin{equation}\label{nonlinear}
\rho = \nabla\cdot D_{\rm e},
\qquad
J = \nabla\times H-\partial_t D_{\rm e}.
\end{equation}

For this example, \eqref{nonlinear} defines a local assignment of $\rho$ and $J$ to every admissible field datum. The associated field energy density is $u = E \cdot D_e - \mathscr{L}$. With the complete field densities and fluxes 

\begin{gather}
u = \mathscr{L}'E^2-\mathscr{L}, \nonumber \\ 
g=S=\mathscr{L}'E\times B, \nonumber \\
T_{ab} = \mathscr{L}'(E_aE_b+B_aB_b)
-\delta_{ab}(\mathscr{L}+\mathscr{L}'B^2),
\label{quantities}
\end{gather}
momentum and energy conservation, \eqref{momentum} and \eqref{energy}, hold for every smooth field satisfying \eqref{maxwell12}. The sources depend smoothly on the field values and their first derivatives. The energy is normalized by $u(0,0)=0$ and is positive at every nonzero field:
\begin{equation}
u = \frac{E^2 +B^2}{2} + \frac{a X^2}{2}(5E^2 + B^2)>0.
\end{equation}
The construction respects proper rotations and conventional parity and time-reversal assignments. Its source law nevertheless differs from Maxwell's. E.g. $E= k x_1 \hat{e}_1, B =0,$ with constant nonzero field gradient $k$, gives
\begin{equation}
\rho = k \bigg( 1+ \frac{15 a k^4 x_1^4}{4} \bigg), \quad J =0,
\end{equation}
instead of the Maxwellian value $\rho =k$ in these units. \newline
Without the assumption of at most quadratic dependence on fields, smooth sources, positive normalized energy, the stated symmetries, and energy and momentum conservation do not select the linear source laws. Further restrictions are needed. This example does not establish that quadratic dependence is the only sufficient restriction or the weakest one. No global evolution of the nonlinear model is established here. Causal propagation requires further conditions on the Lagrangian \cite{Schellstede2016,RussoTownsend2024Energy}, and the present example does not satisfy them. For source free infinitesimal perturbations about $E=0$ and a constant magnetic field of magnitude $B_0>0$, a plane wave propagating perpendicular to the background, with its magnetic perturbation parallel to the background, has speed $v$ given by
\begin{equation}
v^2 = \frac{4 + 15 a B_0^4}{4 + 3 a B_0^4} > 1.
\end{equation}
Therefore energy positivity does not imply causal propagation.
\subsection{Verification of energy and momentum conservation}

Let us be keeping the units and definitions of the preceding subsection. The differential identity $d \mathscr{L} = D_e \cdot d E - H \cdot dB$ and $u = E\cdot D_e - \mathscr{L}$ give

\begin{gather}
\partial_t u = E\cdot\partial_t D_e + H\cdot\partial_t B, \nonumber \\
\nabla\cdot(E\times H) =H\cdot(\nabla\times E)- E \cdot (\nabla\times H).
\end{gather}

Since $S=E\times H$, adding these equations cancels the $H$ terms by \eqref{maxwell12} and we obtain $E \cdot (\partial_t D_e - \nabla \times H) = - E \cdot J$ which by \eqref{nonlinear} proves the energy conservation \eqref{energy}. \newline
Let us write $D_a \defe (D_e)_a$ for the displacement components distinct from the datum $D$ in \eqref{data}. Rewriting \eqref{quantities} as $T_{ab} = E_a D_b + H_a B_b - \delta_{ab} ( \mathscr{L} + H_i B_i)$ gives
\begin{align}\label{divergence}
\partial_b T_{ab} &= E_a \partial_b D_b + D_b (\partial_b E_a - \partial_a E_b) +B_b(\partial_b H_a - \partial_a H_b) \nonumber \\
&=\rho E_a - (D_e \times (\nabla \times E))_a + [(\nabla \times H)\times B]_a.
\end{align}
Where we have used $\partial_a \mathscr{L}= D_b \partial_a E_b - H_b \partial_a B_b, \; \partial_b B_b =0,$ and $\partial_b D_b = \rho$. Since $g = D_e \times B, $ Faraday's law in \eqref{maxwell12} gives
\begin{equation}\label{derivative}
\partial_t g = (\partial_t D_e) \times B - D_e \times (\nabla \times E).
\end{equation}
Subtracting \eqref{derivative} from \eqref{divergence} cancels $-D_e \times (\nabla \times E)$ and leaves $\rho E + (\nabla \times H - \partial_t D_e)\times B = \rho E + J \times B$. This concludes proving momentum conservation \eqref{momentum} using the nonlinear sources defined by \eqref{nonlinear}.

\end{document}